\documentclass[useAMS,usenatbib]{mn2e}
\usepackage{graphicx,amssym}
\newcommand{\bc}{\begin{center}}
\newcommand{\ec}{\end{center}}

\title[Modelling galaxy mass evolution from $z\sim 0.8$ to today]
      {Modelling galaxy stellar mass evolution from $z\sim 0.8$ to today}
\author[L.Wang, Y.P. Jing]
       {Lan Wang$^{1,2}$$\thanks{Email:
           wanglan@mpa-garching.mpg.de}$,
         Y.P. Jing$^2$
        \\      
        $^1$MPA/SHAO Joint Center for Astrophysical Cosmology,
Shanghai Astronomical Observatory, Nandan Road 80, Shanghai
200030, China\\
$^2$Key Laboratory for Research in Galaxies and Cosmology, Shanghai
Astronomical Observatory, \\
Chinese Academy of Sciences, Nandan Road 80, Shanghai 200030, China}
\begin{document}

\date{Accepted 2009 ???? ??. 
      Received 2009 ???? ??; 
      in original form 2009 ???? ??}

\pagerange{\pageref{firstpage}--\pageref{lastpage}} 
\pubyear{2009}

\maketitle

\label{firstpage}
\begin{abstract}
We apply the empirical method built for redshift $z=0$ in the previous
work of Wang et al. to a higher redshift, to link galaxy stellar mass
directly with its hosting dark matter halo mass at redshift of around
$0.8$. The $M_{stars}$-$M_{infall}$ relation of the galaxy stellar
mass $M_{stars}$ and the host halo mass $M_{infall}$ is constrained by
fitting both the stellar mass function and the correlation functions
at different stellar mass intervals of the VVDS observation, where
$M_{infall}$ is the mass of the hosting halo at the time when the
galaxy was last the central galaxy. We find that for low mass haloes,
their residing central galaxies are less massive at high redshift than
those at low redshift. For high mass haloes, central galaxies in these
haloes at high redshift are a bit more massive than the galaxies at
low redshift. Satellite galaxies are less massive at earlier times,
for any given mass of hosting haloes. Fitting both the SDSS and VVDS
observations simultaneously, we also propose a unified model of the
$M_{stars}$-$M_{infall}$ relation, which describes the evolution of
central galaxy mass as a function of time. The stellar mass of a
satellite galaxy is determined by the same $M_{stars}$-$M_{infall}$
relation of central galaxies at the time when the galaxy is accreted
and becomes a sub-component of a larger group. With these models, we
study the amount of galaxy stellar mass increased from z$\sim 0.8$ to
the present day through galaxy mergers and star formation. Low mass
galaxies ($<3\times 10^{10}h^{-1}M_{\odot}$) gain their stellar masses from
z$\sim 0.8$ to $z=0$ mainly through star formation. For galaxies of
higher mass, we find that the increase of stellar mass solely through
mergers from $z=0.8$ can make the massive galaxies a factor $\sim 2$
larger than observed at $z=0$, unless the satellite stellar mass is
scattered to intra-cluster stars by gravitational tidal stripping or
to the extended halo around the central galaxy that is not counted in
the local observation. We can also predict stellar mass functions of
redshifts up to $z\sim 3$, and the results are consistent with the
latest observations. Future more precise observational data will allow
us to better constrain our model.
\end{abstract}


\section{Introduction}
\label{sec:intro}

To study how galaxies form and evolve in their hosting dark matter
haloes, a lot of efforts have been made to link galaxy properties with
the properties of dark matter haloes which they reside in.  The usual
methods used include galaxy kinematics\citep{erickson1987} and galaxy
lensing\citep{mandelbaum2005,mandelbaum2006}, which measure the mass
of hosting dark matter haloes directly. Semi-analytic models trace the
gas cooling, star formation, and feedback processes `ab initio' to get
the properties of galaxies of present
day\citep{lucia2006,bower2006}. Halo occupation distribution models
study the galaxy-halo connection empirically, to model galaxy
properties using certain assumed formula to describe the galaxy-halo
relation\citep{jing1998,berlind2002,yang2003,vale2004,conroy2006,wang2006}.

For the models that describe the formation and evolution history of
galaxies, the statistics that are commonly used to constrain these
models include: number statistics such as number density, luminosity
function and stellar mass
function\citep{bullock2002,zehavi2005,moster2009}, spatial clustering
properties described by correlation functions\citep{jing1998,
  yang2003, zehavi2005}, void probability
distribution\citep{vale2004}, pairwise velocity
dispersion\citep{jing1998}, the Tully-Fisher relation\citep{yang2003},
the colour distribution of galaxies and the clustering dependence on
galaxy colour\citep{bosch2003, kravtsov2004, zheng2004,wang2007}.

The current models of galaxy-halo connection mainly focus on low
redshift study, particularly of the present day, simply because we can
handle well the observational statistics only at low redshifts.  With
the development of large scale galaxy redshift surveys, observations
are obtained not only for the local Universe, but also toward higher
redshifts. Surveys aiming at studying the properties of high redshift
galaxies include DEEP2 survey\citep{davis2003}, the COMBO-17
survey\citep{wolf2004}, the VIMOS-VLT Deep
Survey(VVDS)\citep{lefevre2005}, the Cosmic Evolution Survey
(COSMOS)\citep{scoville2007} and zCOSMOS survey\citep{lilly2009}. 
With the help of these surveys,
luminosity functions of different types of galaxies are obtained up to
redshift $z>7$\citep{reddy2008,bouwens2008}. Correlation functions in
luminosity bins reaches redshift of $z\sim
1$\citep{coil2006,pollo2006}.  Stellar mass functions have been
detected for galaxies up to redshift of $z\sim 5$\citep{drory2005,
  fontana2006, elsner2008}. Correlation functions in bins of stellar
mass have been studied for galaxies of redshift $z\sim
1$\citep{meneux2008,meneux2009} for VVDS and zCOSMOS observations.

Based on the observational data obtained at high redshifts, several
works have been done to model the properties of galaxies at early
epoch.  Some works use the HOD models to study high z galaxy
properties\citep{cooray2005}, as well as galaxy clustering
properties\citep{bullock2002,yan2003,cooray2006,conroy2006,white2007},
which focus mainly on the clustering dependence of galaxy luminosity.
Most recently, \citet{zheng2007} uses the HOD method to model the
clustering of DEEP2 galaxies as a function of luminosity, which
reaches redshift of $z\sim 1$.  While luminosity is the most studied
and easily got property of a galaxy, stellar mass is nevertheless a more
fundamental property, since luminosity may be affected a lot by dust
attenuation. \citet{moster2009} uses a statistical approach to
determine the relation between galaxy stellar mass and its hosting
dark matter halo, and constrains the evolution of galaxy stellar mass
by fitting the stellar mass functions at different redshifts taken
from \citet{drory2004}.

In the previous work of \citet{wang2006}, an empirical method has been
used to link galaxy stellar mass directly with its hosting dark matter
halo mass. The method falls in between the semi-analytic approach and
the halo occupation distribution approach. Positions of galaxies are
predicted by following the merging histories of halos and the
trajectories of subhaloes in the Millennium
Simulation\citep{springel2005}. The stellar mass of galaxies at
redshift $0$ is related to the quantity $M_{infall}$ by a double power
law function. $M_{infall}$ is defined as the mass of the halo at the
time when the galaxy was last the central dominant object. Parameters
describing the function are constrained by fitting both the stellar
mass function and the correlation functions at different stellar mass
intervals from SDSS observation\citep{li2006a}. The derived
$M_{stars}$-$M_{infall}$ relation is in excellent agreement with the
determination from galaxy-galaxy weak lensing measurement of
\citet{mandelbaum2006}. In this study, we will apply this method to an
earlier epoch. By fitting the statistic results of VVDS observation,
we can study the connection between galaxy mass and its hosting halo
mass at redshift of around $0.8$. Based on the relations obtained both
of today and of higher redshift, we can study the evolution of galaxy
stellar mass from $z\sim 0.8$ to present day.

After a galaxy falls into a larger group and becomes a satellite, its
surrounding gas is shock-heated and star formation ceases in a short
timescale. The stellar mass of satellite galaxies should remain about
the same as the time when they are accreted. In this case, the stellar
mass of a satellite galaxy is determined by the
$M_{stars}$-$M_{infall}$ relation of central galaxies at the time of
infall. This inspires us to describe the $M_{stars}$-$M_{infall}$
relation for all galaxies at any redshift in a uniform way, by
modelling the evolution of $M_{stars}$-$M_{infall}$ relation of
central galaxies. Assuming that satellite stellar mass does not change
after infall, the relation for satellite galaxies follows the relation
of central galaxies at an earlier epoch when it is accreted. We will
explore this unified model in \S 4. With this model, we can also test
if a significant amount of satellite disruption by tidal forces is
required by current observations.

This paper is organised as follows: in Sec.~\ref{sec:model}, we
present the model for fitting VVDS observations to get the relation
between galaxy stellar mass and the hosting halo mass at $z\sim
0.8$. Based on the models describing galaxy stellar masses both at low
and high redshifts, we analyse in Sec.~\ref{sec:massincrease} how
galaxies gain their masses from redshift of $0.8$ to today. In
sec.~\ref{sec:uniform} we build a unified model to fit observations of
both low and high redshifts simultaneously, assuming that satellite
stellar mass is determined by the $M_{stars}$-$M_{infall}$ relation of
central galaxies at the time of its accretion, and study the mass
growth of galaxies from $z=0.8$ based on this model. In
sec.~\ref{sec:SMF} we predict the stellar mass functions of higher
redshifts based on our two best-fit models, and compare our results
with recent observations.  This work is based on the Millennium
Simulation\citep{springel2005}.


\section{model}
\label{sec:model}

\begin{figure*}
\bc
\hspace{-1.4cm}
\resizebox{16cm}{!}{\includegraphics{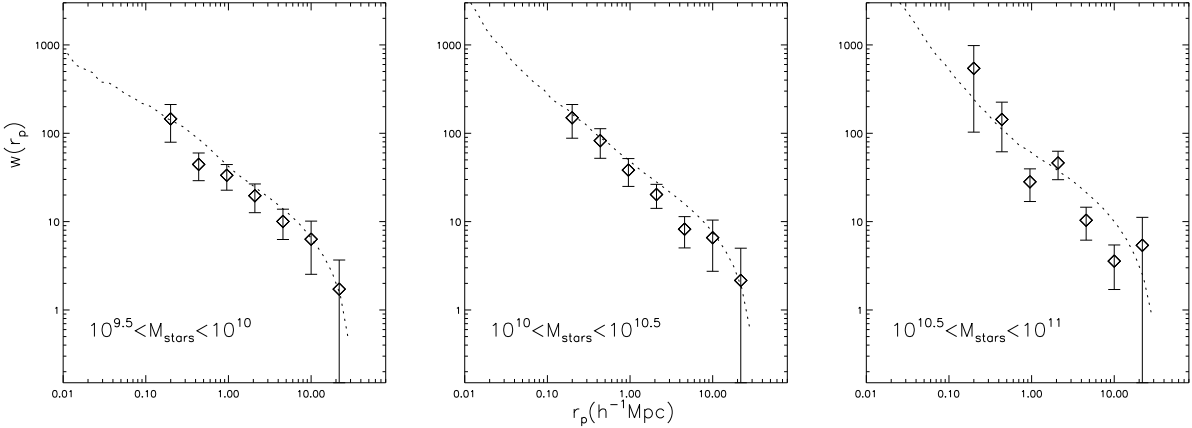}}\\%
\caption{Correlation functions derived by applying the
  $M_{stars}$-$M_{infall}$ relation of z=0\citep{wang2006} to z=0.83, compared with the
  observational results from VVDS of redshift of around
  $0.8$\citep{meneux2008}. Symbols with error bars are from
  observation, while dotted lines represent the model prediction.
}
\label{fig:VVDS}
\ec
\end{figure*}

As mentioned in Sec.~\ref{sec:intro}, in the previous work of
\citet{wang2006}, the relation between the galaxy stellar mass and its
hosting halo mass at infall time have been studied by fitting
both the stellar mass function and correlation functions at different
stellar mass bins from the SDSS observation. This relation can be
described by a double power law form formula. To study the
$M_{stars}$-$M_{infall}$ relation at higher redshifts, as a first test, we
assume that the $M_{stars}$-$M_{infall}$ relation at higher redshifts are
the same as that of present day. We test whether the resulting stellar mass
functions and correlation functions are consistent with the observation at higher
redshifts. We
simply apply the best-fit $M_{stars}$-$M_{infall}$
relation at z=0 of \citet{wang2006} to 
higher redshifts, and derive stellar masses of all galaxies at each time. The
$M_{stars}$-$M_{infall}$ relation 
is described as follow:
\begin{displaymath}
{{M}_{stars}} = \frac{2}{(\frac{{M}_{infall}}{{{M}_{0}}})^{-\alpha}+(\frac{{ M}_{infall}}{{{M}_{0}}})^{-\beta}}{\times}{k}.
\end{displaymath}
The scatter in  $\log(M_{stars})$ at a given value
of $M_{infall}$  was  described with a Gaussian function of a
width $\sigma$. 
The best-fit model to the SDSS observation had the following parameters: $M_0=4.0\times10^{11}h^{-1}M_{\odot}$, 
$\alpha=0.29$, $\beta=2.42$, $\log{k}=10.35$ and $\sigma=0.203$ for  
central galaxies and $M_0=4.32\times10^{11}h^{-1}M_{\odot}$, $\alpha=0.232$, 
$\beta=2.49$, $\log{k}=10.24$ and $\sigma=0.291$ for satellite galaxies.

\begin{figure}
\bc
\hspace{-0.4cm}
\resizebox{8cm}{!}{\includegraphics{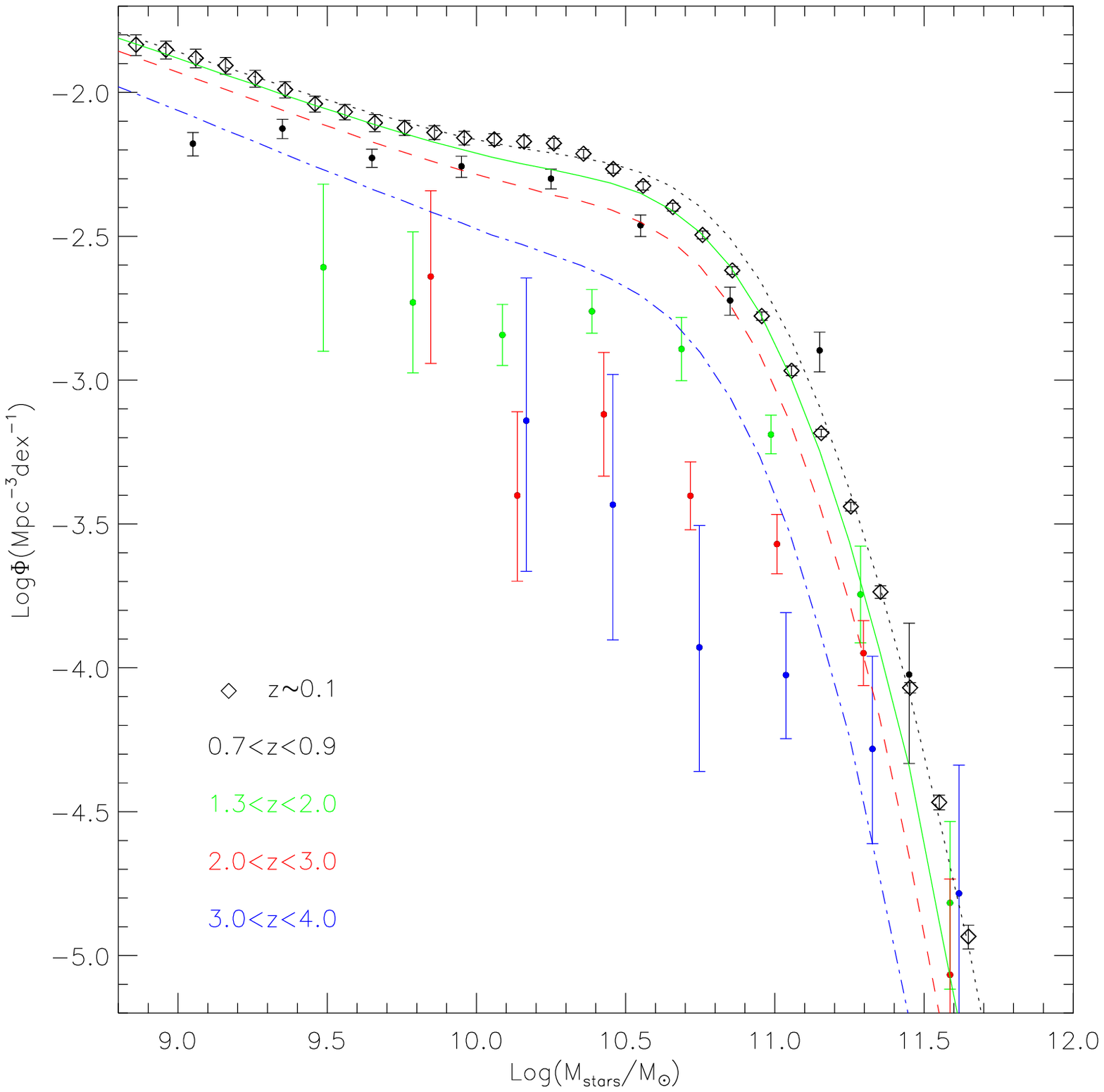}}\\%
\caption{
Galaxy stellar mass functions at different redshifts. Symbols with
error bars are observational results. Black diomends are SDSS
observation of $z\sim0.1$\citep{li2009}. Black points are VVDS results in the
redshift bin of $[0.7,0.9]$ from \citet{pozzetti2007}. Green, red and blue points are results
from \citet{marchesini2008}, in three redshift bins.  Lines are model
prediction, with black, green, red and blue ones corresponding to
results at redshifts of $0.83$, $1.5$, $2.07$ and $3.06$,
respectively. Stellar masses of galaxies
are normalized to the Chabrier IMF\citep{chabrier2003}.
}
\label{fig:SMF}
\ec
\end{figure}

Once we get stellar masses of galaxies at a certain redshift, we can
calculate the stellar mass function and also the clustering results at
different stellar mass bins at that redshift, combined with the
dynamical information of substructures in the simulation. The position
of each galaxy is derived directly from Millennium Simulation by
following the evolution of substructures.  Fig.~\ref{fig:VVDS} shows
the projected correlation functions at three different stellar mass
bins at redshift of around $0.8$. Symbols with error bars are results
from VVDS observation\citep{meneux2008}, and dashed lines are the
derived results from our simple model. The projected correlation
functions predicted by the model are in reasonably good agreement with
the observation from VVDS.

\begin{figure*}
\bc
\hspace{-1.4cm}
\resizebox{16cm}{!}{\includegraphics{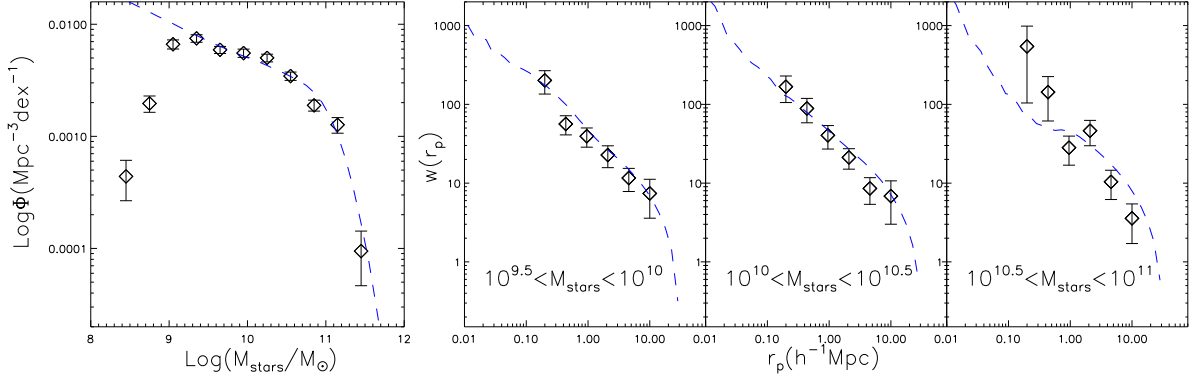}}\\%
\caption{
The best-fit model results for fitting both stellar mass
function\citep{pozzetti2007} and correlation
functions in three different stellar mass bins\citep{meneux2008} from VVDS observation. Symbols are
observational results, and dashed lines are the best
fit model results. 
}
\label{fig:bestVVDS}
\ec
\end{figure*}

Fig.~\ref{fig:SMF} shows the observed stellar mass functions at
different redshifts(symbols), compared with the stellar mass functions
derived from our model(lines), with the $M_{stars}$-$M_{infall}$
relation taken the same as that of present day.  Considering the fact
that different initial mass functions(IMF) were adopted in different
observations, we convert all stellar masses to the case of the Chabrier
IMF\citep{chabrier2003}, in this figure and all the figures of stellar
mass functions hereafter. The galaxy mass obtained with the Salpeter
IMF\citep{salpeter1955} is divided by $1.70$, and that with the
Kroupa\citep{kroupa2001} IMF is divided by $1.104$\citep{cowie2008}.
It is clear from Fig.~\ref{fig:SMF} that the stellar mass functions
are not reproduced under the simple assumption that
$M_{stars}$-$M_{infall}$ relation does not evolve with
time. Observationally, the number of galaxies with low stellar mass
decreases dramatically towards higher redshifts, while the number of
high mass galaxies stays roughly unchanged with different
redshifts. The model derived results, however, show an opposite
trend. The number of low mass galaxies evolves quite slowly, and
remains almost the same till redshift $2$, while the number of high
mass galaxies evolves a lot, with a much smaller number of galaxies
existing at higher redshifts. These results show that the
$M_{stars}$-$M_{infall}$ relation must vary at different redshifts. A
reasonable model should in general give more massive galaxies and
fewer low mass galaxies at higher redshifts than at the local
universe.  As shown in Fig.3 of \citet{moster2009}, the stellar mass
function is more sensitive to the change of model parameters, while
the $\chi^2$ of correlation functions stays flatter around minimum in
a large range of parameter sets. This explains why the correlation
functions can be well reproduced while the stellar mass functions show
such a large discrepancy. Therefore, correlation function alone is not
enough to constrain the $M_{stars}$-$M_{infall}$ relation at $z\sim
0.8$. At the end of this section, we will show that stellar mass
function alone is not enough either to give a good constraint on
the relation, at least in fitting the current observational data
results.


From Fig.~\ref{fig:VVDS} and Fig.~\ref{fig:SMF} we know that we need
to fit both the stellar mass function and the two point correlation
functions to constrain the underlying $M_{stars}$-$M_{infall}$
relation at higher redshifts. Current studies of VVDS observation give
both the stellar mass function\citep{pozzetti2007} and correlation
functions in different stellar mass bins\citep{meneux2008} at redshift
of $\sim0.8$. We therefore focus on constraining $M_{stars}$-$M_{infall}$
relation by these VVDS results, building models based on the
simulation output at redshift of $0.83$. Following the method of
\citet{wang2006}, we assume that the $M_{stars}$-$M_{infall}$ relation
at redshift of $\sim 0.8$ can be described by a double power law
form. The relation is determined by five parameters for central and
satellite galaxies separately, which includes in total 10 parameters
in the modelling. When applying the modelling to VVDS results, we
notice that compared with the observations of the local Universe, the
observational results from higher redshift survey like VVDS provide
much fewer data points. Besides, the error bars of these data points
are still large, which yields weak constraint on the model. Therefore,
on the basis of our previous model of \citet{wang2006}, we alter only
the critical mass(and the k parameter simultaneously) when
constructing the new model, and keep the rest parameters describing
the power law slopes and the relation scatter the same as those for
the local Universe of SDSS. In this case, we now have four free
parameters that are needed to be constrained.

By fitting both the stellar mass function and the correlation
functions of VVDS observation, we get our best-fit model. Stellar mass
function is provided by \citet{pozzetti2007}, in the redshift bin of
[0.7,0.9], with a mean redshift of $0.81$. The correlation functions
are from \citet{meneux2008}, based on the galaxy sample in the
redshift range $z=[0.5,1.2]$, with mean redshift of $0.85$. The best fit is defined as the
one that makes the ${\Xi}$ minimum.                                             
\begin{displaymath}
{\Xi} = \frac{{\chi}^2(\Phi)}{N_{\Phi}}+\frac{{\chi}^2_{corr}}{N_{corr}}
\end{displaymath}
with
\begin{displaymath}
{{\chi}^2(\Phi)}= \sum_{N_{\Phi}}{[\frac{\log{\Phi}-\log{\Phi_{VVDS}}}{\sigma(\log{\Phi_{VVDS}})}]^2}
\end{displaymath}
and
\begin{displaymath}
{{\chi}^2_{corr}}= \sum_{N_{corr}}{[\frac{\log{w(r_p)}-\log{w(r_p)_{VVDS}}}{\sigma(\log{w(r_p)_{VVDS}})}]^2}
\end{displaymath}
$N_{\Phi}=7$, is the number of points over which the stellar mass
function is measured, ranging from $10^{9.5}M_{\odot}$ to
$10^{11.6}M_{\odot}$. $N_{corr}=18$, is the number of points
over which the correlation function is measured, ranging from $0.2$ to
$10.0h^{-1}$Mpc, in three different stellar mass bins.

\begin{figure}
\bc
\hspace{-0.6cm}
\resizebox{7cm}{!}{\includegraphics{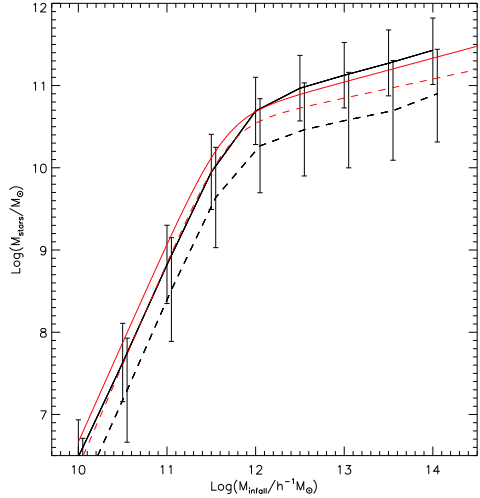}}\\%
\caption{
The best-fit $M_{stars}$-$M_{infall}$ relation at $z=0.83$ of
central galaxies(black solid line) and satellite galaxies(black dashed
line), by
fitting both stellar mass function and correlation functions of VVDS
observation. For comparison, relations of central(red solid line) and satellite
galaxies(red dashed line) from \citet{wang2006} at $z=0$
are also plotted.
}
\label{fig:bestMMinfall}
\ec
\end{figure}

Our best-fit model has the parameters: $M_0=6.31\times10^{11}h^{-1}M_{\odot}$, 
 $\log{k}=10.48$ for  
central galaxies and $M_0=5.03\times10^{11}h^{-1}M_{\odot}$,
$\log{k}=9.99$ for satellite galaxies. The resulting ${\Xi}=3.07591$,
with ${\chi}^2(\Phi)/N_{\Phi}=1.944$. These parameter values are
listed in Tab. 1 to be compared with the best-fit parameters of
modelling SDSS observation\citep{wang2006}.
Fig.~\ref{fig:bestVVDS} shows the best-fit model results. Symbols with error bars are
the VVDS observation, and dashed lines are the model results. The
stellar mass function is well fitted. The clustering is also reasonably 
reproduced. In Fig.~\ref{fig:bestMMinfall}, we plot the derived
best-fit $M_{stars}$-$M_{infall}$ relation at
redshift of $0.83$, for central(black solid line) and satellite(black
dashed line) galaxies respectively. In
comparison, we over-plotted the relations at redshift $0$ in red
lines(solid line for central galaxies and dashed line for satellite galaxies). The result
shows that for a given infall mass of hosting halo, the galaxy mass
changes with redshift, and the dependence on redshift depends on
mass. For massive haloes, the central galaxies inside these haloes are
a bit more massive than the galaxies within the same mass of haloes at
$z=0$. For less massive haloes, the mass
of central galaxies is smaller at higher redshift. For satellite
galaxies, however, the mass
of galaxies is much smaller toward higher redshift at all mass scales. 


 \begin{table*}
 \caption{Best-fit parameter values for the relations between
$M_{infall}$ and $M_{stars}$ in different models to fit SDSS and VVDS observations.}
\begin{center}
 \begin{tabular}{cccccccc} \hline
   &         &  $M_0(h^{-1}M_{\odot})$ & $\alpha$ & $\beta$ & $log(k)$  & $\sigma$& $\tilde{\chi}^2$\\ \hline
   SDSS  & central & 4.0$\times10^{11}$ & 0.29 & 2.42 & 10.35  & 0.203& 5.010\\
Wang et al. 2006         & satellite & 4.32$\times10^{11}$ & 0.232 & 2.49 & 10.24  & 0.291& \\
            \hline
VVDS  & central & 6.31$\times10^{11}$ & 0.29 & 2.42 & 10.48  & 0.203& 3.076\\
    this work        & satellite & 5.03$\times10^{11}$ & 0.232 & 2.49 & 9.99  & 0.291& \\
            \hline
unified model          & z=0 & 3.21$\times10^{11}$ & 0.29 & 2.42 & 10.17  & 0.240& 6.352\\
 this work            & z$ \sim 0.8$ & 4.34$\times10^{11}$ & 0.29 & 2.42 & 10.15  & 0.240& 8.103\\\hline
 \end{tabular}
\end{center}
 \end{table*}

\begin{figure*}
\bc
\hspace{-1.4cm}
\resizebox{16cm}{!}{\includegraphics{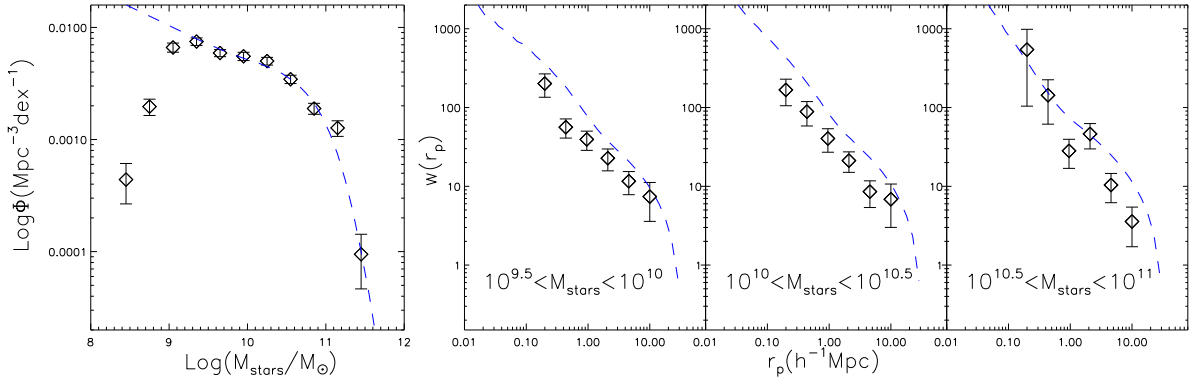}}\\%
\caption{
The best-fit model result when fitting only stellar mass function from VVDS
observation\citep{pozzetti2007}. Symbols are observation results, and
dashed lines are the best-fit model results. 
}
\label{fig:bestSMFfit}
\ec
\end{figure*}

In a recent paper, \citet{moster2009} claim that stellar mass
function alone is enough to constrain the relation between galaxy
stellar mass and its hosting halo mass, since the fit to correlation functions has a much wider
range at its $\chi^2$ minimum than the fit to stellar mass functions.
However, this may not be true for higher
redshift case. In Fig.~\ref{fig:bestSMFfit}, we show the derived
stellar mass function and correlation functions for the best-fit model
when fitting only the stellar mass function of VVDS observation. The
results show that the clustering of galaxies is generally
over-predicted in this case. Therefore, we believe that fitting both
stellar mass function and the correlation functions simultaneously is required
to get reasonable fit, at least for the current observational data we
can get. 

\section{Mass Increase from z$\sim 0.8$ to z$=0$}
\label{sec:massincrease}

From high redshift to the present day, dark matter haloes get larger
through mergers, and galaxies inside them also become bigger in
size. The galaxies gain their masses through either mergers with other
galaxies, or by forming new stars. Using the model we build in
Sec.~\ref{sec:model} we already know the galaxy masses at redshift of
$0.83$. We also have galaxy stellar masses of today according to the
model built at $z=0$ from \citet{wang2006}. The galaxy mass of today
is a total amount of stellar component of galaxy stellar mass that
already exists at redshift of $0.83$, the mass increase resulting from
mergers with other galaxies, and in addition the newly formed stars
during the time interval. By tracing the merger histories of
haloes/subhaloes and hence the galaxies that reside in these
haloes/subhaloes, the amount of stellar mass that was added through
mergers can be calculated. Combined with the stellar mass of galaxies
at both $z=0.83$ and $z=0$, the stars that were newly formed during
the time interval between these two redshift epochs can be predicted.
  
\begin{figure}
\bc
\hspace{-0.6cm}
\resizebox{6.5cm}{!}{\includegraphics{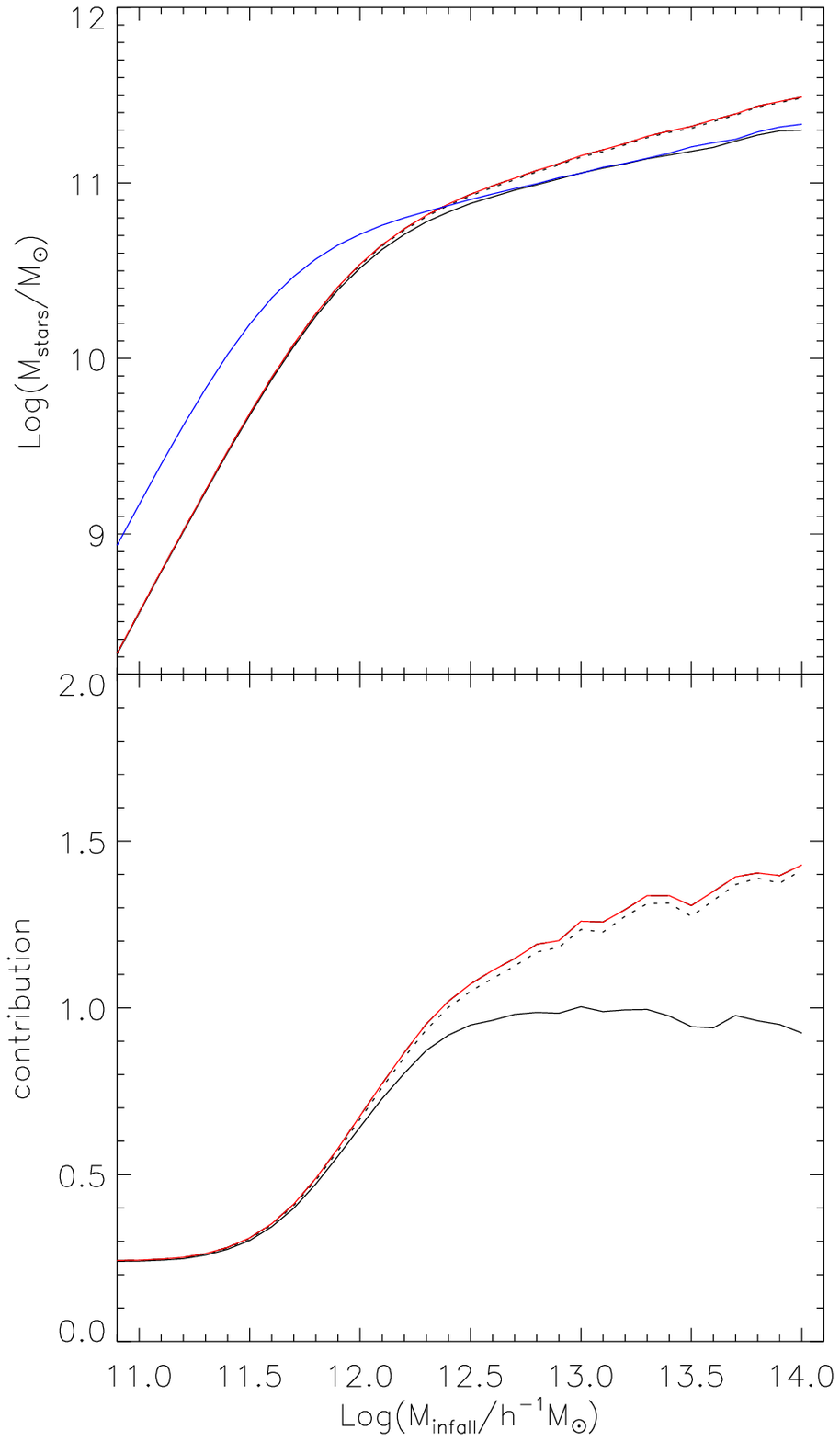}}\\%
\caption{
Galaxy mass increase from $z=0.83$ to $z=0$, for different hosting
halo masses. Upper panel: for a given mass of haloes at present day,
black solid line shows the median value of
the most massive progenitors of the central galaxies that reside in them. 
Dotted black line is the total amount of mass of both the most massive
progenitor and the merged galaxies which were satellites
at $z=0.83$. Red solid line represents the total mass of the most
massive progenitor at $z\sim 0.8$ and all merged galaxies till
$z=0$. Blue line is the central galaxy mass of present day, which represents
$M_{stars}$-$M_{infall}$ relation at $z=0$. Bottom panel:
corresponding ratio of each mass to the galaxy mass of present day.
}
\label{fig:massincrease}
\ec
\end{figure}

For a galaxy that resides in a halo of given mass at $z=0$, we trace
back through merger trees to its most massive progenitor at
$z=0.83$. We plot in Fig.~\ref{fig:massincrease} the median relation
between the stellar mass of the most massive progenitor at $z=0.83$ of
a galaxy and the mass of its host halo at present day in black solid
line. Among the galaxies that merge into this most massive progenitor,
some of them are galaxies that already exist at $z=0.83$, including
both central and satellite galaxies at that time. The other galaxies
are newly formed galaxies after $z=0.83$, and merge into the main
group before the present day. From our fitted model results we know
that the $M_{stars}$-$M_{infall}$ relation evolves with
time. Therefore, at the time of each merger, the mass of the merged
galaxy should not be the same as its mass at the time of redshift
$0.83$. We get the galaxy mass at the time of each merger by
interpolating $M_{stars}$-$M_{infall}$ relation between $z=0.83$ and
$z=0$, assuming that the model parameters evolve linearly with
redshift.

In Fig.~\ref{fig:massincrease}, the dotted black line is the sum of
the stellar masses at $z=0.83$ of the most massive progenitor and of
its satellites merged in. The Red solid line is the result when the
merged mass from central galaxies is added, including both the stellar
mass existing at $z=0.83$ and those newly formed since then. The
contribution from the merged central galaxies is much smaller compared
with the mass from merged satellite galaxies. Blue lines are the mass
of galaxies of present day, according to our model fit result for SDSS
observation\citep{wang2006}. In the bottom panel of
Fig.~\ref{fig:massincrease}, the corresponding mass ratio of each
component to the galaxy of the present day is shown.

From Fig.~\ref{fig:massincrease} we can tell that for galaxies that
reside in haloes of mass less than $10^{12}h^{-1}M_{\odot}$, the
mergers since $z=0.83$ contributes to the stellar mass growth by a very
small fraction, which can even be ignored. Compared with the galaxy
mass at present day, their most massive progenitors contribute from
about 20 percent to around 60 percent of the present day mass, while
the rest of the mass of $z=0$ galaxies should come from star formation
of the central galaxy itself. However, for high mass galaxies whose
hosting halo masses are more than $10^{12.5}h^{-1}M_{\odot}$, the
story is totally different. The mass of the most massive progenitor
galaxy at $z=0.83$ is comparable to its present day mass. When taking
into account the stellar component of other merged galaxies, the total
mass is significantly larger than the galaxy mass of the present
day. This paradoxical result is the consequence of hierarchical
merging in the current cosmological model. Here we have adopted the
merger trees constructed by \citet{lucia2006} for galaxies by taking
into account the dynamical time scales. We found that the merged
fraction of stellar mass does not change when the merger time scale of
\citet{jiang2008} is adopted. There are several possibilities to
reconcile the observations at low and high redshifts in the
hierarchical model. One is that satellite galaxies are tidally
disrupted in a significant amount of stellar mass before merged into
the central galaxies \citep{yang2009,wetzel2009}. However, as we see in \S
4, the significant tidal disruption is not strongly required by
current observational data, because the unified model (details in \S
4) in which we assume no tidal disruption matches the observational
data equally well. Another possibility is that central galaxy mass
observationally determined may differ from the galaxy mass defined in
simulation, due to the limit size of the observed central region of a
galaxy. Besides, when counting for the merged mass in simulation, all
stellar masses of the merged galaxies is added, while observationally,
the more extended stellar halo around the central galaxy may not be
fully counted in. Therefore observationally determined galaxy mass can
be smaller than the simulated total mass. With the model advocated
here, we expect to study and discriminate these possibilities when the
model parameters can be determined better with future high redshift
samples of galaxies. Future observations of intra-cluster (group)
stars can also help testing these possible mechanisms.

In any case, qualitatively we should be able to conclude that for low
mass galaxies, they gain quite a fraction of their present day mass
through star formation from $z=0.83$ to today, while for high mass
galaxies, most of their present day mass already exist at redshift of
$0.83$, and star formation is not active for these galaxies. This is
consistent with the result of \citet{wang2007} where they constrain
the star formation histories of galaxies by fitting the spectral
distribution properties of galaxies. This is also consistent with the
well-known fact that low mass galaxies are in general more active and
blue in colour, while high mass galaxies are red and passively
evolved.

In \citet{zheng2007}, they use the HOD method to model the
luminosity-dependent projected correlation function of DEEP2 and SDSS
surveys, and study the stellar mass growth of galaxies, by estimating
the stellar mass from galaxy luminosity and colour based on the mean
relation between these properties. Compared with their result shown in
their Fig.9, the contribution of high redshift galaxies to galaxy mass
of the present day has a similar trend with halo mass, but the
absolute values of our model are in general higher than their
result. Besides, the increase at halo mass of around
$10^{12}h^{-1}M_{\odot}$ is more steeper in our model. Notice that
they are modelling observation of DEEP2, whose redshift range is
around 1, while our model is to fit the VVDS, with redshift of around
$0.8$, this discrepancy is in the right direction to be explained by
the increase of galaxy mass through time.  As pointed out by
\citet{zheng2007}, around $25$ percent more of the stellar mass could
have been in place in the $z\sim1$ progenitors due to the fact that
the DEEP2 sample they used are not entirely volume limited for red
galaxies. This would decrease the discrepancy between these two model
results. However, for massive haloes, galaxies inside them are still
more massive at higher redshift in our model than in their model
result. This may be caused by the fact that the DEEP2 sample could
have missed a quite fraction of red massive galaxies because of their
colour selection of target galaxies.

\section{unified model}
\label{sec:uniform}

\begin{figure*}
\bc
\hspace{-1.4cm}
\resizebox{16cm}{!}{\includegraphics{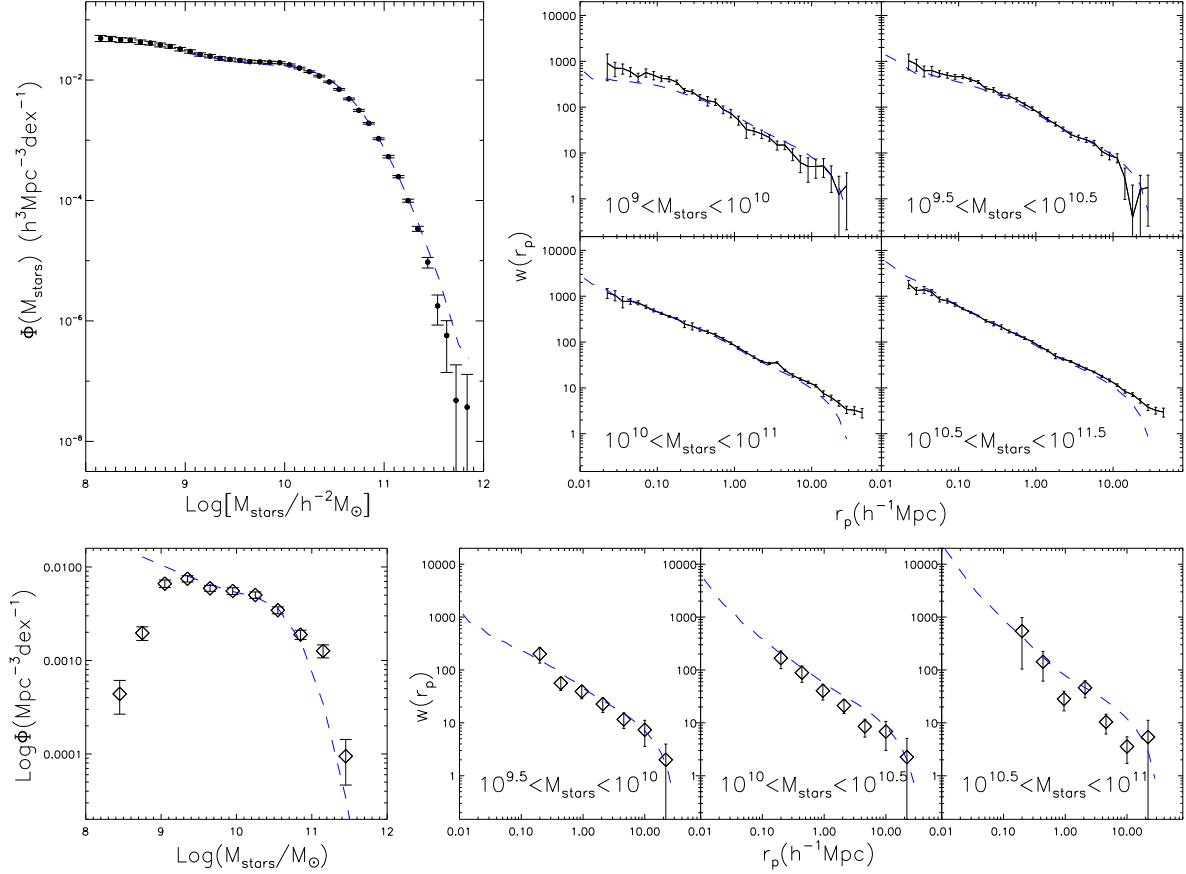}}\\%
\caption{
The best-fit unified model results of stellar mass functions and correlation
functions at $z=0$(upper panels) and $z=0.83$(lower panels),
compared with results from SDSS\citep{li2009,li2006a}(upper panels) and VVDS
observation\citep{pozzetti2007,meneux2008}(lower panels). Symbols with error bars
are from observation, and dashed lines are from our best-fit unified model.
}
\label{fig:unibest}
\ec
\end{figure*}

\begin{figure*}
\bc
\hspace{-1.4cm}
\resizebox{16cm}{!}{\includegraphics{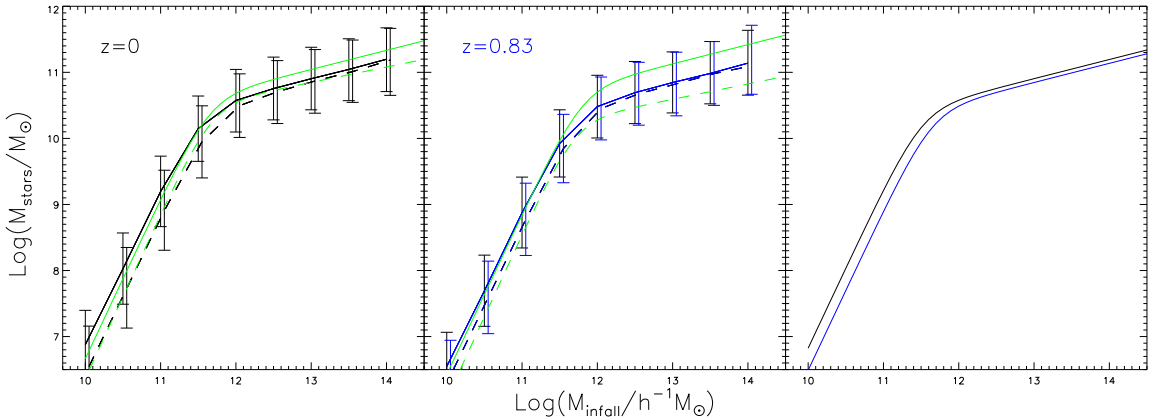}}\\%
\caption{
The $M_{stars}$-$M_{infall}$ relation for the best-fit unified model
at $z= 0$(left panel, black lines) and at $z=0.83$(middle panel,
blue lines). Solid lines are the relation for central galaxies and
dashed lines are results for satellite galaxies. In the left panel, the
$M_{stars}$-$M_{infall}$ relations from the model at $z=0$ of
\citet{wang2006} are over-plotted in solid green lines for central galaxies
and dashed green lines for satellite galaxies. In the middle panel,
solid and dashed green lines are relations for central and satellite
galaxies from our best-fit model in
Sec.2 when fitting VVDS observations only. The right panel compares the median relations for central
galaxies at $z=0.83$(blue line) and at $z=0$(black line). 
}
\label{fig:uniMMinfall}
\ec
\end{figure*}


When a central galaxy falls into a larger group and becomes a
satellite, the gas around the galaxy is shock-heated. The gas stops
cooling and the star formation ceases in a short time scale. The
stellar mass of the galaxy can increase by only a small amount through
star formation. On the other hand, the galaxy stellar mass may
decrease a bit due to the tidal stripping effect. We now assume that
in total, the stellar components of satellite galaxies do not change
compared with the mass at the time when they are central objects. In
this case, $M_{stars}$-$M_{infall}$ relation for a satellite galaxy is
the same as the $M_{stars}$-$M_{infall}$ relation for centrals at an
earlier epoch, when the satellite galaxy falls into a larger group.

We now try to build a unified model to describe the evolution of
$M_{stars}$-$M_{infall}$ relation with redshift, for both central and
satellite galaxies. We assume that at all redshifts, the median of the
$M_{stars}$-$M_{infall}$ relation can be described by a double power
law form. We assume that the power law indexes and the scatter around
the median relation are fixed with time, while the critical mass and
the corresponding normalization parameter evolve with time
linearly. At any given redshift, the stellar mass of central galaxy is
connected to its host halo mass according to the relation. For
satellite galaxy, its stellar mass is connected to its halo mass at
infall time, according to the $M_{stars}$-$M_{infall}$ relation at the
time when the galaxy falls into a larger group and becomes a
satellite.

Based on this picture, we can get the stellar masses for all galaxies at any
given redshift, and calculate their stellar mass functions and
correlation functions. To get the best fit parameters that describe
the evolution relation, we fit at the same time both the observation from SDSS at local
Universe\citep{li2009,li2006a}, and the VVDS results at redshift of
around $0.8$\citep{pozzetti2007, meneux2008}. For simplicity, we fix
the slopes describing the $M_{stars}$-$M_{infall}$ relation to be the
same as the slopes of the relation of central galaxies when fitting
SDSS data only\citep{wang2006}, i.e., $\alpha=0.29$ and $\beta=2.42$. 
In total, we now have
5 parameters to fully describe the $M_{stars}$-$M_{infall}$ relation:
critical mass $M_0$ and $k_0$
for redshift 0, $M_{0.8}$, $k_{0.8}$ to describe the relation at
$z=0.83$, and the scatter $\sigma$ around the median relation. 

The best-fit model is determined when the resulting  $\tilde{\Xi}$ gets its
minimum.                                             
\begin{displaymath}
{\tilde{\Xi}}=
\frac{{\chi}^2(\Phi_{0})}{N_{\Phi,SDSS}}+\frac{{\chi}^2_{corr,0}}{N_{corr,SDSS}}+
\frac{{\chi}^2(\Phi_{0.8})}{N_{\Phi,VVDS}}+ \frac{{\chi}^2_{corr,0.8}}{N_{corr,VVDS}}
\end{displaymath}
with
\begin{displaymath}
{{\chi}^2(\Phi_{0/0.8})}= \sum_{N_{\Phi,SDSS/VVDS}}{[\frac{\Phi_{0/0.8}-\Phi_{SDSS/VVDS}}{\sigma(\Phi_{SDSS/VVDS})}]^2}
\end{displaymath}
and
\begin{displaymath}
{{\chi}^2_{corr,0/0.8}}= \sum_{N_{corr,0.1/0.8}}{[\frac{w(r_p)_{0/0.8}-w(r_p)_{SDSS/VVDS}}{\sigma(w(r_p)_{SDSS/VVDS})}]^2}
\end{displaymath}
$N_{\Phi,VVDS}=7$, $N_{\Phi,SDSS}=29$ are the numbers of points over which the stellar mass
functions are measured in two observations. $N_{corr,VVDS}=6\times 3$,
$N_{corr,SDSS}=30\times5$, are the numbers of points
over which the correlation functions are measured in different stellar mass bins. 

Our best-fit model has the parameters: 
$M_0=3.21\times10^{11}h^{-1}M_{\odot}$, $\log{k_0}=10.17$,
$M_{0.8}=4.34\times10^{11}h^{-1}M_{\odot}$, $\log{k_{0.8}}=10.15$, and
$\sigma=0.240$. The
resulting $\tilde{\Xi}=14.455$, with
${\chi}^2(\Phi_{0})/N_{\Phi,SDSS}=3.133$,
${\chi}^2_{corr,0}/N_{corr,SDSS}=3.218$,
${\chi}^2(\Phi_{0.8})/N_{\Phi,VVDS}=3.778$, and
${\chi}^2_{corr,0.8}/N_{corr,VVDS}=4.325$. The parameters are also
listed in Table.1.
Fig.~\ref{fig:unibest} shows the best-fit model results, compared with the observation from
SDSS and VVDS surveys. All the observations are reasonably
reproduced, although the resulting ${\chi}^2$ is larger, and the fit
is a bit poorer than the
previous model in Sec.2. The constrained evolution of critical mass $M$ and
normalization parameter $k$ can be described as:
\begin{displaymath}
M(t)=\frac{t_0-t}{t_0-t_{0.8}}{\times}({M_{0.8}}-{M_0}) + M_0
\end{displaymath}
\begin{displaymath}
k(t)=\frac{t_0-t}{t_0-t_{0.8}}{\times}({k_{0.8}}-{k_0}) + k_0
\end{displaymath}
$t$ is the age of the universe at different
redshifts. $t_0=13.6098Gyr$, and $t_{0.8}=6.7531Gyr$, are the age of
the universe at redshift 0 and $0.83$ respectively.

Fig.~\ref{fig:uniMMinfall} shows the best fit
$M_{stars}$-$M_{infall}$ relation at two different redshifts. For comparison, we
also over-plot the model result when fitting SDSS and VVDS data separately in green lines. We find that 
at both high and low redshifts, central galaxies in massive haloes are less massive
in the unified model than the separate model. The difference of satellite galaxies between the
two models is smaller. Although the best-fit $M_{stars}$-$M_{infall}$
relations are different in our two models, they can both give
reasonable fits within the observational error range. The main
difference of the two models exists for massive galaxies. Considering that the number of
high mass galaxies is small in observation, which causes large error
bars on the data points, especially for VVDS survey,
the statistics of high mass galaxies are not tightly
constrained. Therefore different sets of parameters can both give
reasonable fits to the observational data. The models can be better
constrained only with improved observational data. 

In this best-fit unified model, from Fig.~\ref{fig:uniMMinfall} we
know that both at $z=0$ and $z=0.83$, satellite and
central galaxies have similar relation for massive haloes, while for low
mass haloes, central galaxies are more massive than satellites. In the
most right panel of Fig.~\ref{fig:uniMMinfall}, we compare the
$M_{stars}$-$M_{infall}$  relations for central galaxies at redshift $0$ and
$0.83$, for the best-fit unified model results. The evolution of this relation
is larger for galaxies in low mass haloes than in more massive
haloes. For all masses of haloes,
central galaxies that reside in them are more massive at lower redshift than at higher
redshift.

\begin{figure}
\bc
\hspace{-0.6cm}
\resizebox{6.5cm}{!}{\includegraphics{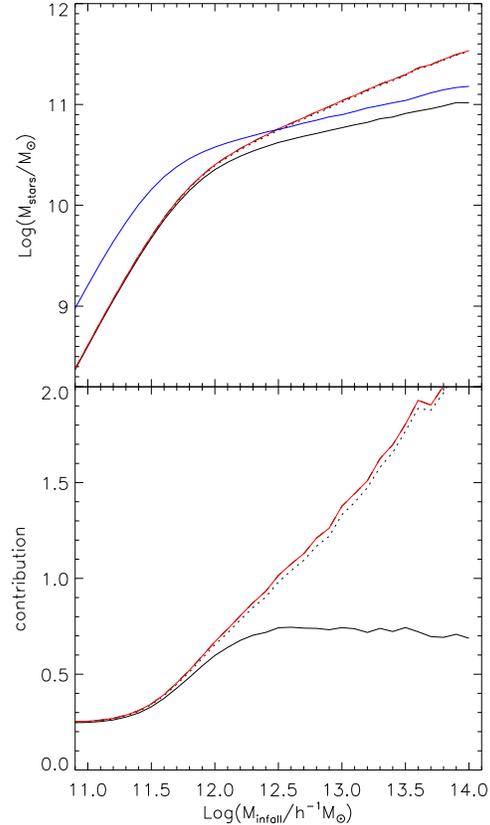}}\\%
\caption{
The same as Fig.~\ref{fig:massincrease} showing galaxy mass increase
from $z=0.83$ to $z=0$ for different host
halo masses, except that for the unified model results. 
}
\label{fig:massincreaseuniform}
\ec
\end{figure}

Based on the method studying the mass increase of galaxies from
redshift of $0.83$ to today described in Sec.~\ref{sec:massincrease}, we can
also detect the mass increase situation according to the result from
the unified model stated above. We perform the same analysis as previously stated, and
show the results in Fig.~\ref{fig:massincreaseuniform}. We find that
the most massive progenitors at $z=0.83$ contribute from around 25
percent at halo mass of $\sim 10^{11}h^{-1}M_{\odot}$ to around 70
percent for haloes massive than $\sim 10^{12.5}h^{-1}M_{\odot}$ to the
galaxy mass of present day. This fraction is much smaller for massive
galaxies than the separate model. The merged fraction of galaxies,
however, is higher in the unified model. Nevertheless, qualitatively
the same as in the separate model, the unified
model also shows that for galaxies in low mass haloes, they gain their
mass from $z=0.83$ to today mainly through star formation, and extra
mechanisms are needed to explain the excess of the total progenitor
masses to the present-day mass for massive galaxies. 
Since in the unified model the tidal tripping is assumed
to be unimportant, the excess of the stellar mass is
presumably located in the stellar haloes of central galaxies.
With upcoming larger deep samples of galaxies, we will consider to
include the tidal stripping effect
in the HOD modeling in a future work.

\section{Stellar mass functions at higher redshifts}
\label{sec:SMF}

In Sec.~\ref{sec:model} and Sec.~\ref{sec:uniform}, we build models
based on the observations of SDSS and VVDS data of galaxy samples
of the local Universe and of redshift of $\sim0.8$. Assuming that our
models can be applied to even earlier epochs, with the parameters evolving
linearly with redshift, we can predict stellar mass functions of
galaxies at higher redshifts. 

\begin{figure*}
\bc
\hspace{-1.4cm}
\resizebox{12cm}{!}{\includegraphics{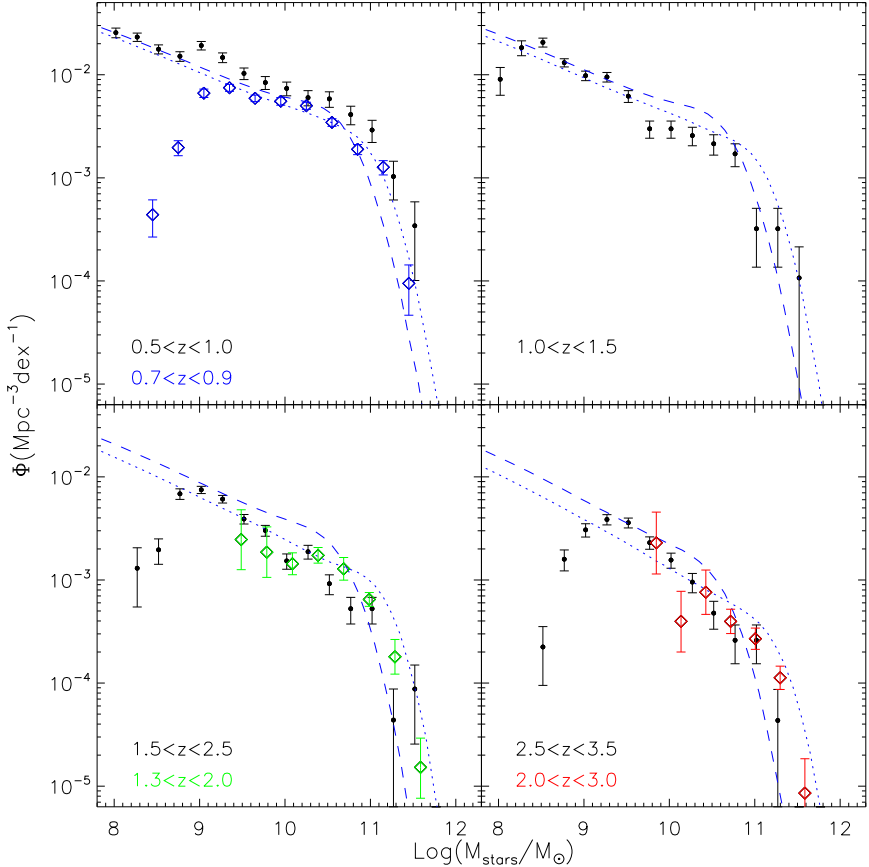}}\\%
\caption{
The prediction of stellar mass functions at high redshift of our
best-fit models. Black points are observations from
\citet{kajisawa2009}, for four redshift bins. Blue dashed lines are
the predictions when applying the $M_{stars}$-$M_{infall}$ relation of
our unified model to higher redshifts of $0.83$, $1.17$, $2.07$, and
$3.06$. Blue dotted lines are predictions of the
best-fit models when SDSS and VVDS data are fitted separately,
assuming parameters evolve linearly with redshift. In the upper left
panel, the observation of \citet{pozzetti2007} is plotted(blue symbols), for $0.7<z<0.9$
galaxy sample, which is used
for building our model at $z0.83$ in Sec.~\ref{sec:model}. In the
lower two panels, observations from \citet{marchesini2008} are
plotted, for $1.3<z<2.0$(green symbols) and $2.0<z<3.0$(red symbols) galaxy
samples. Stellar masses of galaxies
are normalized to the Chabrier IMF\citep{chabrier2003}.
}
\label{fig:SMFmodel}
\ec
\end{figure*}

Fig.~\ref{fig:SMFmodel} presents the results of stellar mass functions
at different redshift bins. Black solid lines are results from recent
observation of \citet{kajisawa2009}, for four redshift
bins. The stellar masses are normalized to the Chabrier
IMF\citep{chabrier2003}. The median redshift of $0.5<z<1.0$, $1.0<z<1.5$, $1.5<z<2.5$ and
$2.5<z<3.5$ galaxy samples are $0.802$, $1.172$, $2.080$, and
$2.952$. In the upper left
panel, also plotted is the
result of \citet{pozzetti2007} of VVDS observation in redshift range
of $(0.7, 0.9)$, which is used in the previous sections to constrain
our best-fit models. In the lower two panels, observations from \citet{marchesini2008} are
over-plotted, for $1.3<z<2.0$(green symbols) and $2.0<z<3.0$(red symbols) galaxy
samples. Blue dotted and dashed lines are predictions from best-fit
models presented in Sec.~\ref{sec:model} and Sec.~\ref{sec:uniform},
at redshifts of $0.83$, $1.17$, $2.07$, and $3.06$. The results from different observations are consistent with
each other, and the predictions from both of our models are also
consistent with these observations. 

Notice that at all redshifts, the prediction from
the unified model gives a higher amplitude of stellar mass function at low
mass end, and a lower amplitude at high mass end, compared with the
prediction from the model where SDSS and VVDS data are fitted
separately. In the unified model, galaxies have smaller masses at high
mass end of haloes, and the satellite galaxy mass is higher at high redshift
than the separate model. These together cause the
difference of amplitudes of stellar mass
functions at low and high mass end. As discussed in Sec.4, although the predictions from our two
models differ at all redshift bins due to the obvious difference of
the $M_{stars}$-$M_{infall}$ relations obtained in the models,
it is hard to tell which model is
a better description of the real universe observed, due to the poor
statistics of current data. Observations with
smaller errors will be able to help us build a more precise model.

\section{Conclusions and discussions}
\label{sec:conclusion}

We apply the empirical model of \citet{wang2006} to higher redshift, to
link galaxy stellar mass with its hosting dark matter halo
mass at redshift of around $0.8$. The 
$M_{stars}$-$M_{infall}$ relation is constrained by fitting both the 
stellar mass function and the correlation functions at different
stellar mass intervals from VVDS observation. We find that the
difference of $M_{stars}$-$M_{infall}$ relation between central and
satellite galaxies at $z\sim 0.8$ are much larger than the galaxies at
present day. At all mass scales of haloes, satellite galaxies at $z=
0.8$ have much lower stellar masses than satellites within the same
mass haloes at $z=0$, while the mass of central galaxies changes less
from high redshift to today. Central galaxies are
less massive in low mass haloes, and more massive in high mass haloes 
at $z=0.8$ than the galaxies at $z=0$.

Under the assumption that the mass of satellite galaxy remains about
the same as it is a central before it infalls to a larger group, we
build a unified model for the $M_{stars}$-$M_{infall}$ relation, which
describes the evolution of galaxy mass as a function of halo mass at
any given redshift. Satellite stellar mass is determined by the
$M_{stars}$-$M_{infall}$ relation of central galaxy at the time of its
infall. The best-fit model of both SDSS and VVDS stellar mass
functions and clustering functions gives an obvious evolution of
$M_{stars}$-$M_{infall}$ relation from $z=0.8$ to $z= 0$, with the
mass of galaxies at higher redshift being lower than the galaxy mass
at the present day, for all masses of hosting haloes. The
$M_{stars}$-$M_{infall}$ relation provided by the unified model is
different as the relations in the models when SDSS and VVDS data are
fitted separately, especially for galaxies with massive haloes and at
high redshift. The central galaxies are less massive and satellite
galaxies are more massive in the unified model than the separate
model. Different sets of parameters can both give reasonable fits to
the observed data, because the statistics of high mass galaxies are
not tightly constrained in observation at high redshift.

We study the amount of galaxy stellar mass growth from $z\sim0.8$ to
today, in either way of galaxy merger or star formation. For the
models when SDSS and VVDS data are fitted separately, we find that for
galaxies that reside in haloes of mass less than
$10^{12}h^{-1}M_{\odot}$, the masses of their most massive progenitors
at $z=0.83$ vary from about 20 percent to 60 percent of the present
day mass. Meanwhile, galaxy mergers contribute only a small fraction
to the galaxy mass of today. This indicates that a large fraction of
$z=0$ stellar masses comes from star formation during the period
between $z=0.83$ and $z=0$. For galaxies within massive haloes, the
total amount of stellar mass from the main progenitor at $z=0.83$ and
from that increased from mergers with other galaxies actually exceeds
the present-day galaxy mass. Although there could be a difference of
the definitions for stellar mass of central galaxies in observation
and in the models based on simulation (for example, the stars in the
envelope of the central galaxies may not be counted in observation),
this indicates that there may be little star formation ongoing in
these massive galaxies. The unified model basically tells the same
story, except that the contribution of the most massive progenitors at
$z=0.83$ to the mass of the present day is no more than 75 percent
even for the most massive galaxies.

Based on the models we build, we give predictions of higher redshift
stellar mass functions, with redshift up to $z\sim 3$. It is
encouraging that our predictions are consistent with recent
observations of stellar mass functions presented by
\citet{marchesini2008} and \citet{kajisawa2009}. However, the
amplitude of the stellar mass functions predicted by the unified model
is always higher at low mass end, and lower at high mass end, compared
with the prediction of the model that fits SDSS and VVDS data
separately. Besides, in both of our models, we assume that for the
$M_{stars}$-$M_{infall}$ relation, its slopes at high and low mass end
and scatter around the median value do not change with time, which may
not actually be true. The current models can be tested by future
observational data, and a more accurate model can be constructed with
better data.

Besides the observational data of VVDS survey used to constrain our
model in this work, galaxy stellar mass function and correlation functions 
in different stellar mass bins are also obtained for zCOSMOS survey by 
\citet{pozzetti2009} and \citet{meneux2009}. As pointed out in \citet{meneux2009},
the zCOSMOS field is centred on an extreme overdensity region. 
We therefore choose to use the VVDS results in this modelling work, although
it is possible that the VVDS field is a bit under-dense on the other
hand. Hopefully this cosmic variance effect can be significantly reduced with 
upcoming larger samples of galaxies, based on which the models can be constrained 
much more tightly.

Recent study of \citet{moster2009} determined the galaxy stellar mass
- halo mass connection at high redshift constrained with stellar mass
functions only. They claim that stellar mass function alone is enough
to constrain the relation between galaxy stellar mass and their
hosting halo mass. We have shown at the end of Sec.~\ref{sec:model}
that this may not be true for high redshift situation, although this
was proved viable when considering models at the local
Universe. However, their result shows the same trend as our
$M_{stars}$-$M_{infall}$ relation in the separate model, although the
quantitatively value of the relation is somewhat different. For low
mass haloes, galaxies at higher redshifts have less stellar mass than
galaxies at a lower redshift. For high mass haloes, galaxies at high
redshift is a bit more massive than galaxies at low redshift. In our
unified model, on the other hand, galaxy mass is always smaller at
high redshift than that at low redshift, but the difference for
galaxies within massive haloes is quite small. The precision of all
these models can be tested only by improved observational results both
on stellar mass functions and on the clustering properties at high
redshifts.

\section*{Acknowledgements}
We thank the referee for detailed suggestions on improving the paper. 
We acknowledge Baptiste Meneux, Masaru Kajisawa and Lucia Pozzetti for 
providing and explaining the
data points of their papers. We are grateful to Guinevere Kauffmann for
helpful discussions.  L.~W acknowledges the financial support of the
joint postdoc program between Chinese Academy of Sciences and the Max
Planck Society. This work is supported by NSFC (10533030, 10821302,
10873028, 10878001), by the Knowledge Innovation Program of CAS (No.
KJCX2-YW-T05), and by 973 Program (No.2007CB815402).

The simulation used in this paper was carried out as part of the
programme of the Virgo Consortium on the Regatta supercomputer of the
Computing Centre of the Max--Planck--Society in Garching. The halo
data together with the galaxy data from two semi-analytic galaxy
formation models is publically available at
http://www.mpa-garching.mpg.de/milleannium/.

\bsp
\label{lastpage}

\bibliographystyle{mn2e}
\bibliography{highz}

\end{document}